\documentclass[nofootinbib,twocolumn]{revtex4}

\usepackage{graphicx}
\usepackage{hyperref}
\usepackage{verbatim}
\usepackage{array}
\usepackage{xcolor}
\usepackage{enumitem}

\begin{document}

\title{Scientific Computing, High-Performance Computing and Data
  Science in Higher Education}
\author{Marcelo Ponce}\email{mponce@scinet.utoronto.ca}
\author{Erik Spence}\email{ejspence@scinet.utoronto.ca}
\author{Daniel Gruner}\email{dgruner@scinet.utoronto.ca}
\author{Ramses van Zon}\email{rzon@scinet.utoronto.ca}
\affiliation{SciNet HPC Consortium, University of Toronto, 256 McCaul Street, Toronto, Ontario M1W 1T5, Canada}
\date{\today}


\begin{abstract}
  We present an overview of current academic curricula for Scientific
  Computing, High-Performance Computing and Data Science. After a
  survey of current academic and non-academic programs across the
  globe, we focus on Canadian programs and specifically on the
  education program of the SciNet HPC Consortium, using its detailed
  enrollment and course statistics for the past four to five
  years. Not only do these data display a steady and rapid increase in
  the demand for research-computing instruction, they also show a
  clear shift from traditional (high performance) computing to
  data-oriented methods. It is argued that this growing demand
  warrants specialized research computing degrees. The possible
  curricula of such degrees are described next, taking existing
  programs as an example, and adding SciNet's experiences of student
  desires as well as trends in advanced research computing.
\end{abstract}

\maketitle

\tableofcontents

\section{Introduction}
\label{sec:intro}

The computational resources available to scientists and engineers have
never been greater.  The ability to conduct simulations and analyses
on thousands of low-latency-connected computer processors has opened
up a world of computational research which was previously
inaccessible.  Researchers using these resources rely on
scientific-computing and high-performance-computing techniques; a good
understanding of computational science is no longer optional for
researchers in a variety of fields, ranging from bioinformatics to
astrophysics.

Similarly, the advent of the internet has resulted in a paradigm where
information can be more easily captured, transmitted, stored, and
accessed than ever before.  Researchers, both in academia and industry
\cite{Radermacher:2013:GIE:2445196.2445351}, have been actively
developing technologies and approaches for dealing with data of
previously-unimaginable scale.  Researchers' ability to analyze data
has never been greater, and many branches of science are actively
using these newly-developed techniques.

Unfortunately, the skills needed to harness these computational and
data-empowered resources are not systematically taught in university
courses \cite{Richards:2011:TRN:2016741.2016801}.  Some researchers, postdocs and students may find
non-academic programs to fill this void, but others either do not have
access to these courses or cannot commit the time to follow them.
These researchers typically end up learning by trial and error, or by
self-teaching, which is rarely optimal.

A number of academic programs that aim to address this issue have
emerged at universities across the world (a few examples are
\cite{Tel-Zur:2014:PEB:2690854.2690857,Burkhart:2014:THC:2690854.2690859}).
Some of these grew out of the training efforts of High Performance
Computing (HPC) centres and organizations ({\it
  e.g}.\ \cite{Stewart:1995:HUC:224170.224209}).  Recognizing the need
for additional skills in their users, computing centres such as those
in the \href{https://www.xsede.org/}{XSEDE} partnership
\cite{XSEDE2014} in the U.S., \href{http://www.prace-ri.eu/}{PRACE} in
Europe, and \href{https://www.computecanada.ca/}{Compute/Calcul
  Canada} have been providing local and online HPC training as part of
their user support.  Universities have also developed graduate
programs in both Scientific and High-Performance Computing, to train
scientists and engineers in the use of these computational resources.

A more-recent complement to these graduate programs is the development
of the degree in Data Science (DS), that is, degrees which focus on
the analysis of data, especially at scale.  These degrees come in a
variety of forms, from multi-year academic graduate programs to
specialized private-sector training.  These programs are in strong
demand at present, as large companies have discovered the value in
thoroughly analysing the vast quantities of customer data which they
collect.  It is expected that this field will continue to grow, and
academic programs will continue to be introduced to meet this demand.


The SciNet HPC Consortium \cite{1742-6596-256-1-012026,
  website:SciNet} is the high-performance-computing consortium of the
University of Toronto.  SciNet provides both computational resources
and specialized user support for Canadian academic researchers, and as
members of its support team, we are responsible for training
researchers, postdocs and graduate students at the University of
Toronto in HPC techniques.  In this paper we give a review of the
current state of graduate-level Scientific Computing, High-Performance
Computing and Data Science academic programs, and endeavour to design
an ideal HPC and DS graduate program.  The paper is organized as
follows.  In Sec.~\ref{sec:HPC-research-review} we discuss how
computation has become an essential ingredient in many academic
research endeavours; in Sec.~\ref{sec:HPC-review} we review the
current status of education in the areas of High-Performance and
Scientific Computing.  In Sec.~\ref{sec:DS-review} we present the Data
Science education efforts at the academic and non-academic level.
Sec.~\ref{sec:futurePrograms} describes what HPC and DS Master's
programs could look like.  We conclude with final remarks and
perspectives for the future in Sec.~\ref{sec:concl}.



\section{The role of HPC in current research}
\label{sec:HPC-research-review}


The breadth of science being as large as it is, it is essentially
impossible to give an overview of the uses of computational methods in
current scientific research.  We will nonetheless attempt a review of
at least some computational scientific research, since the way
computers are used in research (and other realms of inquiry)
influences what should be taught to students.

Astrophysical computational research inherently involves large scale
computing, such as the simulation of gravitational systems with many
particles, magnetohydrodynamic systems, and bodies involving general
relativity.  Atmospheric physics requires large weather and climate
models with many components to be simulated in a variety of scenarios.
High-energy particle physics projects, such as the ATLAS project at
CERN, require the analysis of many recorded events from large
experiments, while other high-energy physics projects have a need for
large scale simulations ({\it e.g.}\ lattice QCD investigations).
Condensed matter physics, quantum chemistry and material science
projects must often numerically solve quantum mechanical problems in
one approximation or another; the approximations make the calculations
feasible but still rely on large computing resources.  Soft condensed
matter and chemical biophysics research often involve molecular
dynamics or Monte Carlo simulations, and frequently require sampling a
large parameter space.  Engineering projects can involve optimizing or
analyzing complex airflow or combustion, leading to large fluid
dynamics calculations.  Bioinformatics often involves vast quantities
of genomic input data to be compared or assembled, requiring many
small computations. Research in other data-driven sciences such as
social science, humanities, health care and biomedical science
\cite{Ihantola:2015:EDM:2858796.2858798}, is also starting to outgrow
the capacity of individual workstations and standard tools in their
respective fields.

Examining these cases in more detail, one can distinguish different
ways in which research relies on computational resources:
\begin{enumerate}
  \item Research that is inherently computational, {\it i.e.}\ it
    cannot reasonably be done without a computer, but which requires
    relatively minor resources ({\it e.g.}\ a single workstation).
  \item Research that investigates problems that do not fit on a
    single computer, and therefore rely on multiple computing nodes
    attached through a low-latency network.
  \item Research that requires many relatively small computations.
  \item Research that requires access to a large amount of storage,
    but not necessarily a lot of other resources.
  \item Research that requires access to a lot of storage, on which
    many relatively small calculations are performed.
\end{enumerate}
The distinction between the various types of research determines the
appropriate systems and tools to use.  Graduate students that are just
starting their research often do not have enough knowledge to make the
distinction (as nobody has taught them about this), let alone select
and ask for the resources that they will need \cite{Richards:2011:TRN:2016741.2016801}.

Note that all five categories fall under ``Advanced Research
Computing'' (ARC). The categories are not mutually exclusive, but
research of the second and third kind are usually associated with HPC,
while the fourth and fifth, and sometimes the first, are associated
with Data Science (DS).  Although there is a lot of overlap between
HPC and DS, these fields require somewhat different techniques.  For
that reason, we will consider separate programs for HPC and DS.

\section{Programs in High-Performance Computing}
\label{sec:HPC-review}

Much of the research presented in the previous section falls in the
category \textit{Scientific Computing} (SC). The growth in the
computational approach to research, both academic and industrial, has
prompted some institutions to develop graduate-level programs designed
to teach the skills needed to design, program, debug and run such
calculations.  These programs, having been in development for more
than two decades, are now fairly wide spread and mature, and are known
by the names ``Scientific Computing'' or ``Computational Science and
Engineering''.  Scientific Computing graduate degrees are offered
internationally in several graduate education hubs around the world
(U.S., England, Germany, Switzerland, {\it etc.}, --- lists of which
can be found at the
\href{https://www.siam.org/students/resources/cse_programs.php}{SIAM}
and \href{http://hpcuniversity.org/students/graduatePrograms}{HPC
  University} \cite{Lathrop:2013:HUG:2484762.2484771} websites).
Canada is no exception here either, with at least eight universities
offering graduate-level programs in Computational Science.
These programs include one-year and two-year Master's programs, as
well as Ph.D.\ programs.  Most of these programs ({\it e.g.}\ the ones
shown in Tables \ref{table:HPCProgram} and \ref{table:BioProgram})
require a final thesis.  The projects and theses are faculty-guided
research projects and are usually one-term long, though, as with all
research, these projects sometimes take longer.

A typical curriculum for a two-year Master's program in Scientific
Computing (in this case from San Diego State University) is presented
in Table~\ref{table:HPCProgram}.  It clearly shows that Scientific
Computing has its roots in research in the physical sciences; the
programs heavily emphasize numerical analysis and scientific
modelling.  In some ways this is not surprising: computers are very
apt at solving such problems, and the formalism of the physical
sciences often lends itself easily to computer programming.  Other
topics of study which are also often encountered in these programs
include finite element analysis, matrix computations, optimization,
stochastic methods, differential equations and stability.
\begin{table}[t]\centering
\begin{tabular}{lcc}\hline
	Course Name	& Type	\\
	\hline\hline
	Introduction to Computational Science & required\\
	Computational Methods for Scientists  & required\\
	Computational Modelling for Scientists& required\\
	Computational Imaging & required\\
	Scientific Computing  & required\\
	Applied Mathematics for Computational Scientists & required\\
	Seminar Problems in Computational Science & required\\
	Computational and Applied Statistics & elective\\
	Computational Database Fundamentals & elective\\
	Research & required\\
	Thesis & required\\
\hline
\end{tabular}
\caption{The curriculum for the two-year
  \href{http://www.csrc.sdsu.edu/masters.html}{Master's program} at
  the Computational Science Research Center at San Diego State
  University \cite{website:SDSUmasters}; this forms a good example of a
  typical Scientific Computing graduate program.
}
\label{table:HPCProgram}
\end{table}

In contrast to Scientific Computing, HPC requires somewhat wider
knowledge; its practitioners need to understand more than just the
theoretical and numerical principles.  They require skills such as
serial and parallel programming (often in several languages, and on
different platforms) and scripting, as well familiarity with numerics,
data handling, statistics, and supercomputers and their technical
bottlenecks.  In addition, these practitioners are usually not
computer science students, so they must cope without that background.
This is somewhat unavoidable as they need to have sufficient domain
knowledge as well. Much of the same holds for Data Science.


\subsection{Academic HPC Programs}
\label{section:HPCprograms}

There are not many academic programs that focus on HPC.  Part of the
reason may be that such programs require access to a
high-performance-computing machine so that students can develop their
skills on real hardware, in a real supercomputing environment.  These
machines require multiple computing nodes which are connected by a
low-latency network.  Fortunately, such systems do not need to be
local: as long as the machine is accessible through the internet the
machine could be used for teaching.  Nonetheless, having the hardware
local to the students lends advantages, since most of the
administrators and analysts of the system are typically available to
assist students with optimizing their codes and developing good
computational strategies.  Not surprisingly, the majority of the
currently offered HPC graduate programs seem to have been developed in
conjunction with or by supercomputer centres.

As examples of High Performance Computing programs, the University of
Edinburgh (UK) offers an
\href{https://www.epcc.ed.ac.uk/msc}{MSc in  High Performance Computing},
  the Universitat Polit\`ecnica de Catalunya/BarcelonaTech (Spain) offers a
\href{http://masters.fib.upc.edu/masters/master-high-performance-computing}{Master in High Performance Computing}
 and a
\href{http://masters.fib.upc.edu/masters/master-data-mining-and-business-intelligence}{Master program in Data Mining and Business Intelligence},
SISSA/ICTP in Italy offers a
\href{http://www.mhpc.it/}{Master in High Performance Computing},
while a collaboration between the University ITMO
(Russia) and the University of Amsterdam (Netherlands) offers a
\href{http://en.hpc-magistr.escience.ifmo.ru/magistr/9/}{Double-Degree Master Programs in Applied Mathematics and Informatics (Computational Science)}.
Note that many of these programs emerged
from locations with a very strong tradition and consolidated
background in HPC.



\subsection{SciNet's HPC Programs}
\label{section:nonacadHPCprograms}

Many HPC centers provide training for their users to fill the
computational-skills gap for the wider scientific community,
such as,
\href{http://www.sdsc.edu/}{SDSC}, \href{http://www.psc.edu/}{PSC}, \href{https://www.tacc.utexas.edu}{TACC}, \href{http://www.ncsa.illinois.edu/}{NCSA},
\href{https://www.bsc.es/}{BSC}, \href{https://www.epcc.ed.ac.uk/}{EPCC}, \href{http://www.cscs.ch}{CSCS},
\href{https://www.sharcnet.ca/}{SHARCNET},
\href{http://www.ace-net.ca/}{AceNet},
\href{http://calculquebec.ca/en/}{Calcul Qu\'ebec},
among many others.
In its capacity as an HPC centre based at the University of Toronto,
SciNet has developed several education and training classes
\cite{website:SciNet-Edu} aimed at helping students and users obtain
the skills and knowledge required to get the most out of
advanced-research-computing resources.  SciNet's training events and
courses are currently taken by researchers, postdocs, and graduate
students across many different departments and even from outside of
the University of Toronto (UofT).  Some of these courses are
considered part of the official curricula and count as graduate level
courses within the Ph.D.\ programs at UofT.

Initially SciNet provided training specifically oriented toward
Scientific Computing, with the purpose of maximizing user
productivity.  These early classes focused on parallel programming
(MPI and OpenMP), best coding practices, debugging, and other
scientific computing needs.  Over the years the breadth of classes has
grown, with classes offered in Linux shell programming, parallel
input/output, advanced C++ and Fortran coding, accelerator
programming, and visualization.  This is in addition to the annual HPC
Summer School which SciNet runs in collaboration with two
  other HPC centres within Compute Ontario\cite{website:CO}, i.e.,
  CAC\cite{website:cac} and SHARCNET\cite{website:sharcnet}.  This
summer school is a week-long intensive workshop on HPC topics, and
more recently, also Data Science topics\footnote{Similar initiatives
  and trends are being carried on by the International HPC Summer
  School \cite{website:iHPCss} within the theme of HPC Challenges in
  Computational Sciences.}.

\begin{figure}[b]\centering
	\includegraphics[width=\columnwidth]{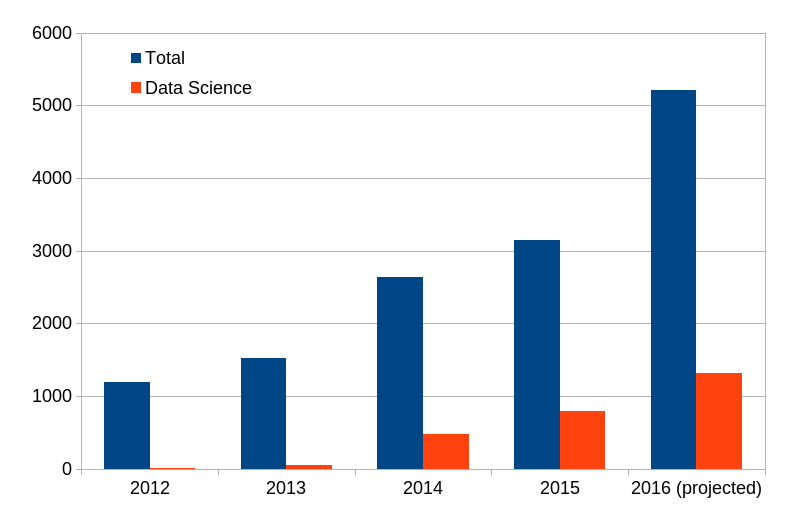}
\caption{Attendance hours at SciNet training and education events, per
  year, for all SciNet classes and Data Science specific classes.}
\label{fig:classhours}
\end{figure}

Table~\ref{table:SciNet-courses} shows the training events and courses
that SciNet has already been teaching in the areas of HPC and Data
Science.
\begin{table*}
\centering
\begin{tabular}{llc}
        \hline
        Course Name 	&	Certificate	&   Credits\\
        \hline\hline
	\textit{Data Analysis with R}$^\ddag$		&	DS/SC	&	12	\\
	Intro to Apache Spark	&	DS	&	3	\\
	Machine Learning Workshop	&	DS/SC	&	6	\\
	Hadoop Workshop			&	DS	&	3	\\
	Scalable Data Analysis Workshop	&	DS	&	12	\\
	Relational Database Basics	&	DS/SC	&	6	\\
	Storage and Input/Output in Large Scale Scientific Projects
					&	DS/SC	&	6	\\
	Workflow Optimization for Large Scale Bioinformatics	&	DS/HPC/CS	&	6	\\
	Python for High Performance Computing
					&	DS/HPC/SC	&	12	\\
	Parallel R			&	DS/HPC/SC	&	3	\\

	Python GUIs with Tkinter	&	DS/SC	&	2	\\

        Scientific Visualization        &       DS/SC   &      6       \\
        Visualizing Data with Paraview          &      DS/SC  &       6       \\

	\hline

	\textit{Scientific Computing for Physicists}$^\ddag$\footnote{This includes 3 previously separate module courses: \emph{Scientific Software Development}, \emph{Numerical Tools for Physical Scientist}, and \emph{High Performance Scientific Computing}.}
					&	HPC/SC	&	36	\\
        \textit{Intro to Research Computing with Python}$^\ddag$ &      HPC/SC  &       12      \\
        Intro to High Performance Computing      &      HPC/SC  &       3       \\
        Intro to Scientific C++  &      HPC/SC  &       6       \\
        Intro to Scientific Programming with Modern FORTRAN      &       HPC/SC	&       7       \\

        Intro to Parallel Programming            &      HPC/SC  &       7       \\
        Programming Clusters with Message Passing Interface	&      HPC/SC  &       12      \\
        Programming Shared Memory Systems with OpenMP		&      HPC/SC  &       6       \\

        Practical Parallel Programming Intensive        &      HPC/SC  &       32      \\

        Intro to GPGPU with CUDA         &      HPC/SC  &       9       \\
        Programming GPUs with CUDA      &      HPC/SC  &       12      \\
        SciNet/CITA CUDA GPU Minicourse         &      HPC/SC  &       12      \\

	Coarray Fortran			&	HPC/SC	&	2	\\
	Parallel I/O			&	HPC/SC	&	6	\\
	
	Debugging, Optimization, Best Practices	&	HPC/SC	&	6	\\
	HPC Best Practices and Optimization	&	HPC/SC	&	3	\\
	HPC Debugging				&	HPC/SC	&	3	\\

	Intro to the Linux Shell		&	HPC/SC	&	2	\\


        Seminars in High Performance Computing  &      HPC/SC  &       4       \\
        Seminars in Scientific Computing        &      HPC/SC  &       4       \\
        \hline\hline
\end{tabular}
\caption{Courses taught by SciNet on
	\emph{Data Science} (\textbf{DS}),
	\emph{High-Performance Computing} (\textbf{HPC}),
	and \emph{Scientific Computing} (\textbf{SC}).
	$\ddag$ denotes courses already recognized by the university as graduate level credits.}
\label{table:SciNet-courses}
\end{table*}
The number and types of classes which SciNet teaches have grown
significantly~\cite{website:SciNet-Training}.  This can be seen in
Figure~\ref{fig:classhours}, which presents the total student
class-hours taught by SciNet over the last four years and the
projection for the current year.  This remarkable growth is a
testament to the latent need for this material to be taught.
The need for this training is supported by the enrolment statistics:
our students constitute 35\% of SciNet's total users, clearly showing
that even in a specialized audience this kind of training is still
needed.

For several years the four-week graduate-style classes offered by
SciNet have been accepted for \emph{graduate} class credit by the
departments of Physics, Chemistry and Astrophysics at UofT.  This was
possible by accepting the classes as ``modular'' (or ``mini'')
courses, one-third semester long, and bundling three such classes into
a full-semester course.  This arrangement has been so popular with
students and faculty that the Physics Department recently listed
SciNet's winter twelve-week HPC class in the course calendar
\cite{website:SciNet-PHY1610H}, allowing graduate students from other
departments in the university to take the class for university credit.

The skills that SciNet aims to transfer are rare and sought-after, and
complement and enhance the skills students learn in regular curricula.
That is why SciNet has developed a set of \emph{Certificate Programs}
\cite{website:SciNet-Certificates}, that users and students can pursue
in \textit{Scientific Computing}, \textit{High Performance Computing},
and/or \textit{Data Science}, once they have completed enough
credit-hours.  As a document that proves the holder has highly
competitive skills, and in lieu of graduate credit for most
  SciNet courses, the certificates are in high demand. In a
resounding endorsement of our teaching, thus far students have
completed a total of 78 certificates (52 in Scientific Computing, 19
in High-Performance Computing and 8 in Data Science).  According to
the current registration and trends, we are projecting to have above a
100 of certificates completed by mid-2016.  Moreover, the feedback
from some of our students was that their SciNet's certificates gave
them an advantage to get jobs in industry and the financial sector.

\section{Programs in Data Science}
\label{sec:DS-review}

The wide adoption of the internet in the professional and the personal
sphere ushered in the age of ``Big Data''.  The ease of recording of
people's online behaviour, and the ability to rapidly move data, lead
to a large, diffuse, complex amount of data waiting to be mined for
useful information.  Because of the typically large size of the data
special hardware and training are often needed.  In contrast to
Scientific Computing and HPC, there are many applications of Data
Science in the private sector, in the medical science, banking,
retail, insurance, and internet industries.

Of these industries, Bioinformatics also has a large component in the
academic world.  Though a more-recent addition to the HPC world, the
bioinformatics field is well-populated with graduate programs, a
testament to its rapid growth and latent demand.  Its emergence as a
major user of HPC systems has resulted in the development of
``Master's of Bioinformatics'', and related degrees.  A typical
Master's program is outlined in Table~\ref{table:BioProgram}, this one
from the Indiana-Purdue University at Indianapolis.  While having many
features in common with a more-standard SC Master's program, such as
the study of programming and algorithms, it exhibits the particular
needs of the bioinformatics community, stressing the importance of
genetics and biological processes, and a lesser emphasis on
mathematics and programming theory.
\begin{table}[t]\centering
\begin{tabular}{lcc}\hline
	Course Name	&	Type	\\
	\hline\hline
	Introduction to Bioinformatics  &required\\
	Seminar in Bioinformatics   & required\\
	Biological Database Management  &required\\
	Programming for Life Science & required\\
	High Throughput Data in Biology  & required\\
	Machine Learning in Bioinformatics  &elective\\
	Computational System Biology & elective\\
	Structural Bioinformatics & elective\\
	Transitional Bioinformatics Applications & elective\\
	Algorithms in Bioinformatics & elective\\
	Statistical Methods in Bioinformatics &elective\\
	Computational Methods for Bioinformatics &elective\\
	Next Generation Genomic Data Analytics & elective\\
	Next Generation Sequencing & elective\\
	Bioinformatics Project & required\\
\hline
\end{tabular}
\caption{The curriculum for the ``Project Track'' two-year
  \href{https://soic.iupui.edu/biohealth/graduate/bioinformatics-masters}{Master's
    of Science in Bioinformatics} at the Indiana University-Purdue
  University in Indianapolis; this forms a good example of a typical
  Bioinformatics graduate program.
}
\label{table:BioProgram}
\end{table}

Degrees in Data Science are relatively new, with the first Master's
program only being introduced in the U.S.\ (by North Carolina State
University) in 2007.  A sample of some of the classes offered in one
such program is given in Table~\ref{table:DataProgram}.  As can be
seen, these programs have a strong focus on data, with statistics,
machine learning, and databases being their standard focus.  Analyzing
data that are too big to fit on a standard desktop computer requires
specialized equipment; such training is also part of these
graduate-level programs, as indicated by the presence of the ``Cloud
Computing'' and ``Distributed Systems'' classes.  Like typical
graduate-level programs, these degrees usually require of the student
a final project or thesis.
\begin{table}[b]\centering
\begin{tabular}{lcc}\hline
	Course Name	& Type	\\
	\hline\hline
	Analysis of Algorithms  & required\\
	Machine Learning & required\\
	Advanced Database Concepts  & required\\
	Distributed Systems &  elective\\
	Advanced Database Concepts &  elective\\
	Cloud Computing &  elective\\
	Information Retrieval & elective\\
	Data Mining &  elective\\
	Web Mining &  elective\\
	Applied Machine Learning &  elective\\
	Complex Networks and Their Applications &  elective\\
	Relational Probabilistic Models &  elective\\
	Internship in Data Science &  elective\\
\hline
\end{tabular}
\caption{A selection of the courses available for the
  \href{http://www.soic.indiana.edu/graduate/degrees/data-science/ms-data-science/index.html}{Master's
    of Data Science} at the Indiana University.
}
\label{table:DataProgram}
\end{table}


One could argue that the novelty of methods in Data Science is due to
its roots in Business Analytics (BA), where the objective is to make a
decision. The field has certainly grown beyond that, and BA is now
considered a sub-field of Data Science.  Another more-recently
developed sub-field is in the realm of health care (``Health
Informatics'').  Because these sub-fields are directly applicable to
the private sector (and the associated revenue streams these present)
these have become the most-commonly implemented post-graduate
programs.  The Business Analytics programs focus on using data to
refine business administration, as well as develop marketing
strategies.  Health Informatics programs concentrate on using clinical
data to optimize health care processes.

The practical focus of Data Science is reflected in the presence of an
internship in the Data Science curriculum listed in
Table~\ref{table:DataProgram}.  Internships in such programs are
similar to other co-op-type arrangements: the student works with an
employer for a semester, allowing the student to gain hands-on
experience applying the skills learnt during such period.

\subsection{Academic Data Science Programs}

Graduate level programs in Data Science are not difficult to find.
For instance, programs in bioinformatics (a data-driven field), can be
found on the
\href{https://www.iscb.org/iscb-degree-certificate-programs}{web site
  of the International Society for Computational Biology}.  It speaks
to the rapid rise of the field bioinformatics, that there are more
bioinformatics programs available than Scientific Computing programs.
Examples lists of other Data Science programs can be found
\href{http://analytics.ncsu.edu/?page_id=4184}{at the NCSU analytics
  web site},
\href{http://www.predictiveanalyticstoday.com/online-business-analytics-programs}{the
  online business analytics programs site of
  {\tt predictiveanalyticstoday.com}} and
\href{http://www.onlinecoursereport.com/the-50-best-masters-in-data-science}{at
  {\tt online.coursereport.com}}.  They are not as common as programs in
Scientific Computing, due to the fact that Data Science is relatively
new field of study.  Among those programs about half are offered in
the fields of Business Analytics and Health Informatics, with the
other half being Data Science programs proper.

\subsection{Non-academic Data Science training}

The demand for Data Science skills (or ``Data Analytics'' skills as
they are often called in the private sector) is so high
\cite{Radermacher:2013:GIE:2445196.2445351} that the private sector
has developed programs to meet the growing demand.  A list of such
companies can be found
\href{http://www.skilledup.com/articles/list-data-science-bootcamps}{on
  skilledup.com, which contains a list of data science boot-camps}.  The format of these classes is varied, though they
are all oriented toward a ``boot-camp'' format: some are in person,
some online; some are one-week long, others twelve weeks.  These
programs are very applied, often with one-on-one mentorship with a
seasoned Data Analytics expert.  They also include direct contact with
possible future employers.


Moreover, a great number of these training programs are not focused on
developing analytical thinking or problem-solving skills,
\cite{bootcampsnogood} but rather are aimed at graduated Ph.D.s and
postdocs, whose problem-solving skills are assumed to have already
developed.  This allows them to focus on the technical training
relevant to the job market.  Some of these programs are free, some of
them offer fellowships, and many of them charge on the order of 10-30
thousand US-dollars for a training period of, typically, three months.
These programs have acquired such a level of popularity among young
and recent graduates that the companies offering these programs have
started to perform evaluation tests in order to assess which
candidates are more suitable to be accepted to their programs.
Perhaps the most appealing part for trainees is the networking
platform offered by these programs, as in most of the cases they
provide the opportunity to interact with actual companies looking for
new talent and avoid recruitment layers.

Institutions in the non-profit arena are also starting to offer
programs on Data Science.  For instance the Fields Institute, a
traditional institution for mathematical research, has offered several
workshops and courses, and developed a thematic program on Big Data.
Other examples include the International Centre for Theoretical
Physics (ICTP) and the International School for Advanced Studies
(SISSA), prestigious institutions with a well known tradition in
theoretical physics, now offering training in ``Research Data
Science''.

HPC centers are also venturing into Data Science training, offering
workshops on R, Hadoop, machine learning, {\it etc}.  SciNet started
offering classes with greater data-oriented content ({\it
  cf}. Table~\ref{table:SciNet-courses}) in 2013, with a four-week
class in scientific analysis using Python.  Having now finished its
third year, the class remains popular, with about twenty students
taking the class each year.  The 2015 fall semester also inaugurated
SciNet's first ``Data Science with R'' class, a class focusing on data
analysis techniques using the R language.  This class was very popular
with over twenty-five students finishing the class, and most students
requesting a second installment with more advanced material.
Continuing its growth in the Data Science area, in the last year
SciNet has held workshops in machine learning, scalable data analysis,
and Apache Spark.

\begin{figure}
  \begin{center}
	\includegraphics[width=0.49\columnwidth]{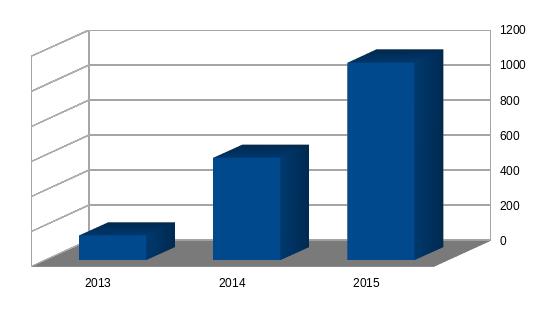}
	\includegraphics[width=0.49\columnwidth]{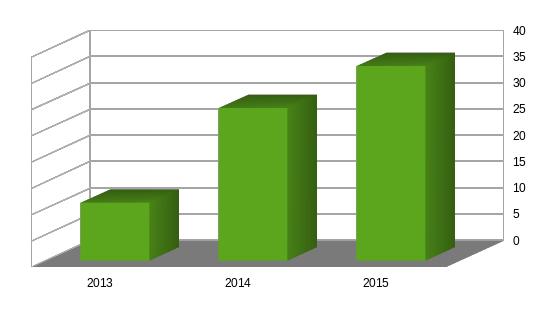}
\caption{Total student hours (left) and taught hours (right) per year,
  for SciNet's Data Science related courses.}
\label{fig:DSclasses}
\end{center}
\end{figure}

Comparing the student- and taught-hours per year shown in
Fig.~\ref{fig:DSclasses}, one sees that the Data Science classes have
been growing consistently, both absolutely as well as relatively (Data
Science related courses roughly constituted less than $2\%$ (2012),
$4\%$ (2013), $12\%$ (2014), $15\%$ (2015), and we project around
$31\%$ (2016) of the total classes taught respectively in each year.



\section{Designing Master's programs in HPC and Data Science}
\label{sec:futurePrograms}

As mentioned above, scientific computing is used by scientists and
engineers as never before, and graduate-level programs in Scientific
Computing are numerous in Canada and around the world.  In contrast,
the development of HPC and Data Science programs is in its early
stages, both in academia and the private sector.  These programs are
being developed to meet the continued shortfall in skill in these
areas, with the McKinsey Global Institute estimating
that the United States will be short 140,000 to 190,000 data analytics
professionals by 2018 \cite{McKinsey_BigData}.

One may wonder whether online learning could not satisfy this need.  A
few examples of MOOCs (Massively Open Online Courses) in HPC and Data
Science do exist.  However, seeing the growth in enrolment in SciNet's
in-person courses and the summer school over the years ({\it
  cf}.~Figs.~\ref{fig:classhours} and \ref{fig:DSclasses}) shows that
many students still prefer the face-to-face format.


Similarly, one may wonder why certificate programs do not suffice
for HPC and DS education. As successful as these programs are, they
have a few disadvantages. Firstly, they are mostly
collections of fairly specific technical training: this leaves no
room for more fundamental material. Secondly, it is also hard to encorporate
an internship or thesis into such a certificate. Finally,
certificates tend to carry less weight than degrees, and, in line
with this, the demand for for-credit courses is larger than that for
not-for-credit courses, as our experience with SciNet's Scientific
Computing graduate course has shown.

A degree program in HPC or DS could offer more academic and
fundamental education, which would leave the student with the
analytical skills and high-level knowledge to stay on top of their
field regardless of changes in computational technology.

In the following sections we propose a curriculum for graduate-level
HPC and DS programs.  One will notice a substantial overlap of topics
with the training courses currently taught by SciNet ({\it cf}. Table
\ref{table:SciNet-courses}).  This is no coincidence: the training
program was developed on the basis of student feedback and requests,
and was a primary inspiration for the curricula proposed here.  The
design was also influenced by the few existing examples of such
programs, as described above.

It should also be emphasized that the programs in HPC and DS are both
designed to allow students to a follow more
industrial/practically-oriented track or an academic/research-oriented
track, by selecting the appropriate set of elective courses and the
corresponding research project/internship.  Additional \emph{advanced}
courses could be made available according to the interest and demand
of the students.

\subsection{Design of an HPC Master's Program}
  \label{sec:SciNet-HPCprogram}

In this section we present a comprehensive and complete curriculum for
a two-year \emph{Master's Program in High-Performance Computing}. 
Students would complete a total of twelve courses.  The five
required courses, each one-term long, set the basis of HPC and ARC
knowledge (including topics such as modern and professional software
development, parallel techniques, performance and optimization, best
practices, distributed systems and resources).  The seven elective
courses allow the student to specialize in a particular area.  In
addition, a final internship or research project would be carried out.

As in any typical Master's program, a student entering the program
will be expected to possess a Bachelor's degree (B.Sc.\ or B.A.).  It
is desirable that students have some background in sciences and the
basics of coding and programming ({\it e.g}.\ Fortran, C, C++);
otherwise it is strongly recommended that students take introductory
programming classes.  Note that courses from the \emph{Data Science}
Master's Program are also eligible to be taken, with consent of the
graduate coordinator or adviser.

The course work for the High-Performance Computing program could
consist of the following courses.

\begin{description}
\item[Software Development (*):] The principles of creating
  modern, maintainable code.
  Special attention is given to designing modular code, and tackling
  scientific computational projects.  Languages: C++/C/Fortran.

\item[Best Practices (*):] Introduction and discussion of techniques
  and methods to be considered when designing and implementing
  computational research projects.  Topics include: version control,
  modularity, libraries.

\item[Performance \& Optimization (*):] Principles and tools for
  measuring performance, finding bottlenecks, and optimizing existing
  code.

\item[Basics of Parallel Programming (*):]  A review of homogeneous and
  heterogeneous architectures is presented, followed by parallel programming
  paradigms such as OpenMP, MPI, and hybrid implementations.


\item[HPC Algorithms:] A review of commonly used algorithms in
  computational science, such as Monte Carlo, implicit and explicit
  methods to solve differential equations, timestepping techniques,
  finite-element and finite-volume methods.

\item[Machine Learning:] Theory and practice of constructing
  algorithms that create models from data.  Topics: probabilistic
  foundations, linear and logistic regression, neural networks,
  Bayesian networks, tree models, support vector machines, density
  estimation, accuracy estimation, normalization, model selection.

\item[Numerical Methods:] A review of the commonly used numerical
  methods in scientific computing, such as linear solvers, fast
  Fourier transforms.  In contrast to the HPC Algorithms, it will
  cover the basic principles and the theory behind the methods only
  briefly, and focus mostly on the implementation and utilization via
  libraries and specific examples.

\item[Scientific Computational Modeling:] This course will offer an
  introduction to the basics of Scientific Computing and a review of
  the most common algorithms and packages used in different fields of
  research computing and computational sciences (astrophysics,
  chemistry, genetics, {\it etc.}  Introduction to computing modeling,
  such as implementation of complex networks based on relational data
  sets, and the evaluation of network properties using graph theory
  elements, among many others.
%

\item[Programming Accelerators:] Some computational problems can be
  computed much more efficiently on accelerators such as graphics
  cards. This course will present an introduction to the use of
  hardware accelerators (GPUs, FPGAs, many-core systems) in HPC.
  Topics will include a review of the hardware, architectures, and
  programming languages, such as, CUDA, OpenACC, and OpenCL.

\item[Research Data Management:] Design strategies, storage management
  and I/O patterns, in order to prevent bottlenecks in massively
  data-driven projects.  Real use cases from various fields
  (bioinformatics, molecular biophysics, medical physics,
  biochemistry, quantum-chemistry, geophysics, {\it etc}.)  have shown that,
  quite often, approaches that work on a desktop do not perform on a
  larger scale.
  A review of the latest policies regarding scientific data
  availability will also be discussed and presented.

\item[Visualization for Scientific Computing:] A review of basic
  visualization concepts and methods with
  applications to scientific data.

\item[Operating System Environment:] Scientific and high performance
  computing is intimately linked to the hardware, OS, and application
  framework on which they used.  This course will help students become
  comfortable working on *nix systems.  Topics such as the command
  line, shell scripting and advanced OS topics will be covered.

\item[HPC Hardware and System Administration:]
  An integrated view on the 
  technology in HPC, the machines, hardware, network, file systems, 
  and what is involved in getting an HPC system up and running.

\item[Student HPC/ARC Seminar:] Weekly sessions running throughout the
  year, with students presenting and discussing current papers and
  research in the fields of HPC and ARC.  Researchers and instructors
  will providing guidance and supervision during the sessions.


\end{description}

In addition to this course work, the program would include an 4-month
internship or independent research project in the final year.

The program presented above is intended to be flexible.  In the first
year of the HPC program, students might take \emph{HPC Algorithms},
\emph{Parallel Programming}, \emph{Software Development},
\emph{Performance and Optimization}, \emph{Best Practices}, and
\emph{Numerical Methods}, as well as attend the HPC/ARC seminar
series.  In year two, they might take
\emph{Research Data Management}, \emph{Scientific Computational Modeling},
\emph{Graph Theory Applications},
\emph{Machine Learning},
\emph{Visualization for Scientific Computing}, and complete the degree
with a Research Project.

Eligible courses from the \emph{Data Science} Master's Program are
also possible to take with previous consent of the graduate
coordinator or adviser.

\subsection{Design of a Data Science Master's Program}
  \label{sec:SciNet-DSprogram}
Here we present a comprehensive curriculum for a two-year
\emph{Master's Program in Data Science}.  As with the HPC program,
students would complete a total of twelve courses.  The five required
courses set the foundation of data analysis knowledge (including
topics such algorithms, databases, statistics, machine learning) and
seven elective courses allow the student to specialize in a particular
area, such as data mining, machine learning, complex systems.  In
addition, a final internship or research project would be carried out,
in order to obtain real-world experience.

As in any typical Master's program, the entry level will be a
bachelor's degree (B.Sc.\ or B.A.).  It is desirable that students
have some background in sciences and the basics of coding and
programming ({\it e.g}.\ Python, R); otherwise it is strongly
recommended that students take introductory programming classes.
Notice that courses from the \emph{High Performance Computing}
Master's Program may be taken with permission of the graduate
coordinator or adviser ({\it e.g}.\ \textit{Operating System
  Environment}).


The fundamentals of data analysis should be at the core of a program
that will produce analysts capable of tackling real-life problems in
Data Science.  Learning theoretical and practical approaches gives
students an advantage in the real world; this program proposes to
combine both in a unique fashion (similar to how most SciNet courses
are structured).  Topics such as statistical analysis, algorithms and
large data sets are at the centre of the proposed program and
constitute the ``core'' (required) courses.  Additionally, the
elective courses allow students to choose a specialization path by
gaining expertise in areas such as: social data mining, machine
learning, and representation of complex interactions.  The course work
for the Data Science program could consist of the following courses:

\begin{description}
\item[Overview of Data Science:] An overview of the field of Data
  Science, covering data-driven problems from several disciplines such
  as, astronomy, bioinformatics, digital humanities, social sciences,
  {\it etc}.

\item[Basics of Programming:] An introduction to programming, coding
  structures, and basic algorithms. This course will focus on
  languages with data-analytic capabilities, such as \emph{Python} and
  \emph{R}.

\item[Data Analysis Algorithms (*):] An overview of the
  major classes of algorithms, including comparison-based algorithms:
  search, sorting, hashing; information extraction algorithms (graphs,
  databases); graph algorithms: spanning trees, shortest paths, depth
  and breath-first search. 

\item[Machine Learning (*):] Theory and practice of constructing
  algorithms that create models from data.  Topics: probabilistic
  foundations, linear and logistic regression, neural networks,
  Bayesian networks, tree models, support vector machines, density
  estimation, accuracy estimation, normalization, model selection.
  Deep learning algorithms.

\item[Database Theory (*):] The mathematical
  foundations of databases, search and query semantics, relational,
  complex, object-oriented and semi-structured database models.

\item[Database Applications:] SQL, relational algebra, index,
  views, constraints; query complexity; data models, including I/O
  model, streaming model, 
  query optimization,
  optimal join algorithm will be presented.

\item[High Dimensional Data Analysis (*):] Theory and methods for
  exploring high-dimensional data will be presented, including linear
  and non-linear dimension reduction, manifold learning; Euclidean
  representation of proximity and network data; clustering,
  statistical pattern recognition.

\item[Applied Statistics (*):] A review of statistical techniques,
  with applications to data analysis problems.  Topics will include
  hypothesis testing, general linear models, generalized linear
  methods, multivariate statistics.

\item[Best Practices:] Introduction and discussion of techniques and
  methods to be considered when designing and implementing
  computational research projects.  Topics include: version control,
  modularity, libraries.

\item[Data Mining:]
  Algorithmic approaches to discovering patterns in
  large data sets will be covered, including data exploration and
  cleaning; association rules, clustering, anomaly detection, and
  classification.
%
Other examples will include applications
  to text and the web: crawling, indexing, ranking and
  filtering algorithms; applications to search, classification and
  recommendation; link analysis, significance tests.
%
  Data mining techniques applied to social media.  Sentiment analysis,
  polarity classification; graph properties of social networks,
  homophily, distance, influence, spectral methods, information
  diffusion, probabilities models.

\item[Data Security and Integrity:] This course will cover technique
  on how to safely protect the integrity and privacy of sensible data,
  standards on data regulations ({\it e.g}.\ health, patient data,
  denomic data), privacy, encryption algorithms.

\item[Search and Classification Algorithms:] Information retrieval
  theory, with applications, will be explored.  Examples of searching
  algorithms, mathematical representations and matrix applications
  will be covered.

\item[Distributed Systems:] Focus on distributed resources and data
  sets across different architectures and systems. Students will learn
  the skills and abilities of dealing with retrieving and screening
  large data sets in \textit{cloud}-type based systems.

\item[Graph Theory Applications:] An introduction to graph theory and
  complex systems and its applications will be offered.  Material will
  include the implementation of complex networks based on relational
  data sets, and the evaluation of network properties.

\item[Visualization Techniques of Unstructured Data sets:] A review of
  basic visualization concepts, with applications to unstructured
  and/or large data sets and complex and dynamical systems, in order
  to gain insights in the data and visually expose potential
  correlations.

\item[Data Science Seminar Series:] Weekly sessions running throughout
  the year, where students will present and discuss current papers and
  research in the fields of data science and analytics.  Researchers
  and instructors will provide guidance and supervision during the
  sessions.
\end{description}

In addition to this course work, the program would include an
4-month internship or research project.

The first year of classes could consist of
 \emph{Overview of Data Science},
 \emph{Data Analysis Algorithms},
 \emph{Database Theory},
 \emph{Applied Statistics} and
 \emph{Visualization Techniques of Unstructured Data Sets},
while the second year could contain courses in
 \emph{High Dimensional Data Analysis},
 \emph{Graph Theory Applications},
 \emph{Data Mining},
 \emph{Search and Classification Algorithms}, and \emph{Distributed Systems}, and
would be completed by an Internship or a Research Project.

As with the HPC/ARC track, eligible courses from the \emph{HPC}
Master's Program are also possible to take with previous consent of
the graduate coordinator or adviser.

\section{Conclusion}
\label{sec:concl}

We have demonstrated the need for programs in higher education in
High-Performance Computing and Data Science.  If the qualitative
evidence of this seems somewhat limited, it should be understood that
existing HPC and DS programs (academic and non-academic) are still
relatively new.  While some such programs are already in existence, in
many cases students must use non-academic options, or teach the
material to themselves. Academic programs would offer the benefit of
not just teaching specific technical skills, but an education in the
fundamentals of HPC and DS and instilling the analytical skills needed
to adapt to an ever-changing technological landscape.

We have reviewed
existing academic and non-academic education programs, in both HPC and
DS.  In light of this review, we presented a design for Master's
programs in HPC and DS, based on these examples and drawing from the
experience and enrollment statistics in not-for-credit training in HPC
and DS by the SciNet HPC Consortium at the University of Toronto.

To get well-founded graduate master's programs off the ground will not be
without challenges.  It will likely involve partnerships and
discussions with other departments and institutes in order to offer a
stronger and multi-disciplinary program.  Existing HPC Centers, which
already operate between different disciplines, can play a fundamental
role in bringing together such programs.

\bibliographystyle{unsrt}
\bibliography{references_v2}

\end{document}